\documentclass{aa}

\begin{document}

\titlerunning{Face-on, stream-fed intermediate polars}

\title {Face-on, stream-fed intermediate polars: an alternative model for 
       RX~J1914.4+2456 and RX~J0806.3+1527}

\authorrunning{Norton, Haswell \& Wynn}

\author{A.J. Norton\inst{1} \and C.A. Haswell\inst{1} \and G.A. Wynn\inst{2}}

\institute{Department of Physics and Astronomy, The Open University, 
	Walton Hall, Milton Keynes MK7 6AA, UK \and
	Department of Physics and Astronomy, University of Leicester,
	Leicester LE1 7RH, UK}

\date{Accepted 2004 February 27.
      Received 2004 February 9; 
      in original form 2003 December 22}

\abstract{
RX~J1914.4+2456 and RX~J0806.3+1527 have been proposed as double degenerate
binaries with orbital periods of 569s and 321s respectively. An alternative
model, in which the periods are related to the spin of a magnetic white dwarf
in an intermediate polar system, has been rejected by other authors. We show
that a face-on, stream-fed intermediate polar model for the two systems is
viable and preferable to the other models. In each case, the X-ray modulation
periods then represent the rotation of the white dwarf in the binary reference
frame. The model explains the fully modulated X-ray pulse profiles, the X-ray
spectra, the antiphase between X-ray and optical/infrared modulation, the lack
of longer period modulation, and the low level of polarization. The optical
spectrum of RX~J0806.3+1527 suggests that Balmer series lines are present,
blended with HeII lines. This is unlike the spectra of any of the known AM CVn
stars and suggests that the system is not a double degenerate binary. The
optical spectrum of RX~J1914.4+2456 has spectral features that are 
consistent with those of a K star, ruling out
the double degenerate models in this case. The lack of optical/infrared
emission lines in RX~J1914.4+2456 may be attributed to a high mass accretion
rate and its face-on orientation. Its reported period decrease may be a short
term spin-up episode driven by the current high $\dot{M}$. Finally we suggest
that there is an observational selection effect such that the face-on
intermediate polars that are detected will all have a stream-fed component, and
the purely stream-fed intermediate polars that are detected will all be face-on
systems.


}

\maketitle

\section{Introduction}

The X-ray sources RX~J1914.4+2456 (hereafter RXJ1914, but now also known as
V407 Vul) and RX~J0806.3+1527 (hereafter RXJ0806) each display a single
coherent X-ray and optical modulation with a period of order several minutes
and no other confirmed modulations in any waveband observed. The most widely
accepted models interpret the periods as due to orbital motion in a double
degenerate binary system.  In this scenario, the two systems have the shortest
orbital periods of any known binary star. A review of both systems is
presented by Cropper, Ramsay \& Wu (2003).

In the rest of this section we summarise the observational history of the two
objects and note the difficulties in reconciling their behaviour with that of
double degenerate binary systems. In section 2 we present simple analytical
estimates of the physical properties of face-on stream-fed intermediate polars
(IPs) and simulations demonstrating the accretion flow in these systems. We
then consider how the various observational characteristics of RXJ1914 and
RXJ0806 may be understood in terms of this model.

\subsection{RX J1914.4+2456}

RXJ1914 was identified from the {\em ROSAT} all sky survey as a 569s X-ray
pulsator and suggested initially to be a member of the then recently recognised
class of soft IPs, with 569s representing the spin period of a white dwarf
(Motch et al 1996).  Subsequent {\em ROSAT} observations failed to reveal any
longer periods, apparently ruling out the IP model because an orbital period
modulation might be expected. The unusual X-ray pulse profile, with zero flux
for half the cycle, was claimed to require a system close to $90^{\circ}$
inclination in conflict with the lack of X-ray eclipses; a model in which the
569s period is the beat period was also ruled out (Cropper et al 1998). 
Instead, a double degenerate polar model was suggested, with a 569s orbital
period -- the first magnetic analogue of the AM CVn stars (Cropper et al 1998).
Optical and infrared spectroscopy and photometry of the optical counterpart
revealed that the V-band through to J-band modulations are all roughly
anti-phased with the X-ray modulation, that the optical and infrared spectra
show no emission lines, and that the system exhibits negligible polarisation
(Ramsay et al 2000; 2002a). As this cast doubt on the double degenerate polar
model, alternative interpretations were suggested including a double degenerate
Algol (direct impact accretor) system (Ramsay et al 2002a; Marsh and Steeghs
2002) and a double degenerate electrically powered system (Wu et al 2002). 
Further doubt regarding the double degenerate accretor models (polar or Algol
type) was raised by the discovery that the X-ray modulation frequency of
RXJ1914 is increasing. If this period decrease represents a secular
evolutionary trend, the observation is in direct conflict with a system which
accretes via Roche lobe overflow from a degenerate star, since in that case a
secular orbital period increase would be expected (Strohmayer 2002). Very
recently, Strohmayer (2004) has claimed that a power spectrum of the {\em
Chandra} X-ray data shows evidence for a sideband structure to the 569s signal.
This indicates a previously unseen longer period in the system of around $\sim
1$ hour. The data also confirm the steady decrease of the 569~s period, with a
frequency derivative of $6 \times 10^{-18}$~Hz~s$^{-1}$. A final piece of
evidence that poses a question for all three double degenerate models was
provided by a spectrum obtained by Danny Steeghs (private communication) which
has spectral features that are consistent with those of a K star.

\subsection{RX J0806.3+1527}

RXJ0806 was discovered as a 321s X-ray pulsator amongst serendipitous X-ray
sources observed by the {\em ROSAT} HRI, and suggested as an IP (Israel et al
1999). Its X-ray modulation is remarkably similar to that of RXJ1914, showing a
50\% duty cycle with the flux reduced to zero between pulses and it too was
suggested to be a double degenerate polar (Burwitz \& Reinsch 2001; Israel et
al 2002).  The faint optical counterpart (Israel et al 1999; Burwitz \& Reinsch
2001) displays an optical period coincident with that seen in X-rays and no
convincing longer period (Ramsay, Hakala \& Cropper 2002b, Israel et al 2002),
although Reinsch (2003) reports some evidence for a possible 4700~s period in
both optical and X-ray data that was also hinted at in the earlier
observations.  As with RXJ1914, RXJ0806 displays antiphased X-ray and optical
pulse profiles (Israel et al 2003a). Unlike RXJ1914, the spectrum of RXJ0806
shows faint, broad emission lines superimposed on a blue thermal continuum
(Israel et al 2002). Further optical spectroscopy by Reinsch (2003) rules out a
main sequence secondary of any spectral type, but allows the possibility that
the system could contain a brown dwarf donor star.  Polarimetric data obtained
by Israel et al (2003b) show that RXJ0806 exhibits no circular polarization,
but that linear polarization is present at a level of 1.7\%$\pm$0.3\%. It has
been reported (Strohmayer 2003; Hakala et al 2003) that RXJ0806 shows a
decreasing period, which if interpreted as a secular change in the orbital 
period would again rule out a system in which material is
accreted from a degenerate companion (double degenerate polar or Algol models).
However Woudt \& Warner (2003) cast doubt on the measurement of this period
derivative claiming that the period count is too uncertain.  

\section{Face-on, stream-fed intermediate polars}

We use our calculations and recent results from the literature to reassess the
proposal that RXJ1914 and RXJ0806 are face-on, stream-fed IPs. This model was
considered, but rejected, for RXJ1914 (Cropper et al 1998; 2003) and for
RXJ0806 (Cropper et al 2003; Israel et al 2003a). However, we show below that
their concerns are unjustified. Briefly, we suggest that 569s and 321s are the
synodic rotation periods of magnetic white dwarfs in IPs. The data further
require that the systems are seen close to face-on (i.e. at a low inclination
angle) and are fed directly by the accretion stream (i.e. no accretion disc is
present). As in other IPs, the magnetic axis of the white dwarf is assumed to
be inclined at $\sim 10^{\circ} - 30^{\circ}$ to the white dwarf rotation axis,
which is perpendicular to the orbital plane.  

An extension of this is to suppose that RXJ1914 and RXJ0806 are {\em
double degenerate} face-on, stream-fed IPs. This might be supported by the
colours of RXJ1914 (Ramsay et al 2000) and the blue spectrum of RXJ0806 (Israel
et al 2002) but suffers from the same problems as the other double degenerate
accretor models, which we discuss below. Instead we suggest that the secondary
stars are very late-type main sequence, or in the case of RXJ0806 possibly a 
brown dwarf.

In the following subsections, we begin by reviewing the general properties of
face-on, stream-fed intermediate polars and discuss the conditions required for
stream-fed accretion to occur.  We then show that the data on RXJ1914 and
RXJ0806 self consistently satisfy the predictions of this model.

\subsection{Stream-fed accretion}

In IPs (for a general review, see Patterson 1994) the magnetospheric radius lies
within the white dwarf's Roche lobe and the white dwarf spin is asynchronous
with the orbit, typically $P_{\rm spin} \sim 10^3$s and $P_{\rm orb} \sim$ a
few hours. Two models for the accretion process have been suggested. (i) In
most IPs, the accretion stream from the L1 point feeds a truncated accretion
disc which is disrupted at the magnetospheric radius where material attaches to
the field lines and follows them towards the white dwarf magnetic poles. The
infalling material takes the form of arc-shaped accretion curtains standing
above the white dwarf surface (Rosen et al 1988).  (ii) In some IPs however,
the accretion flow attaches directly to the field lines, without passing
through a disc. Close to the white dwarf, the accretion flow resulting from
stream-fed accretion is similar to that resulting from disc-fed accretion,
except it is considerably less extended in azimuth (e.g. Hellier \& Beardmore
2002), see Figure 1.  Accretion via a combination of disc and stream is also
possible in so called disc overflow accretion. In all models the accretion flow
undergoes a strong shock close to the white dwarf, below which material settles
onto the surface, releasing X-ray to optical emission. Since the magnetic axis
is generally inclined with respect to the white dwarf spin axis, this gives
rise to the defining characteristic of the class, namely pulsed X-ray (and
usually optical) emission. In disc-fed IPs, this pulsation will be at the white
dwarf spin period, whereas stream-fed IPs will exhibit an X-ray pulsation at
the beat (synodic) period, given by 
\begin{equation} 
1/P_{\rm beat} = 1/P_{\rm spin} - 1/P_{\rm orb} 
\end{equation} 
This is due to the accretion flow flipping from one magnetic pole to the 
other.  Disc-overflow accretion will give give rise to X-ray signals at 
both the spin and beat periods (e.g. Norton et al 1997; Beardmore et al 1998).

\begin{figure*}
\setlength{\unitlength}{1cm}
\begin{picture}(10,5)
\put(0,0){\includegraphics{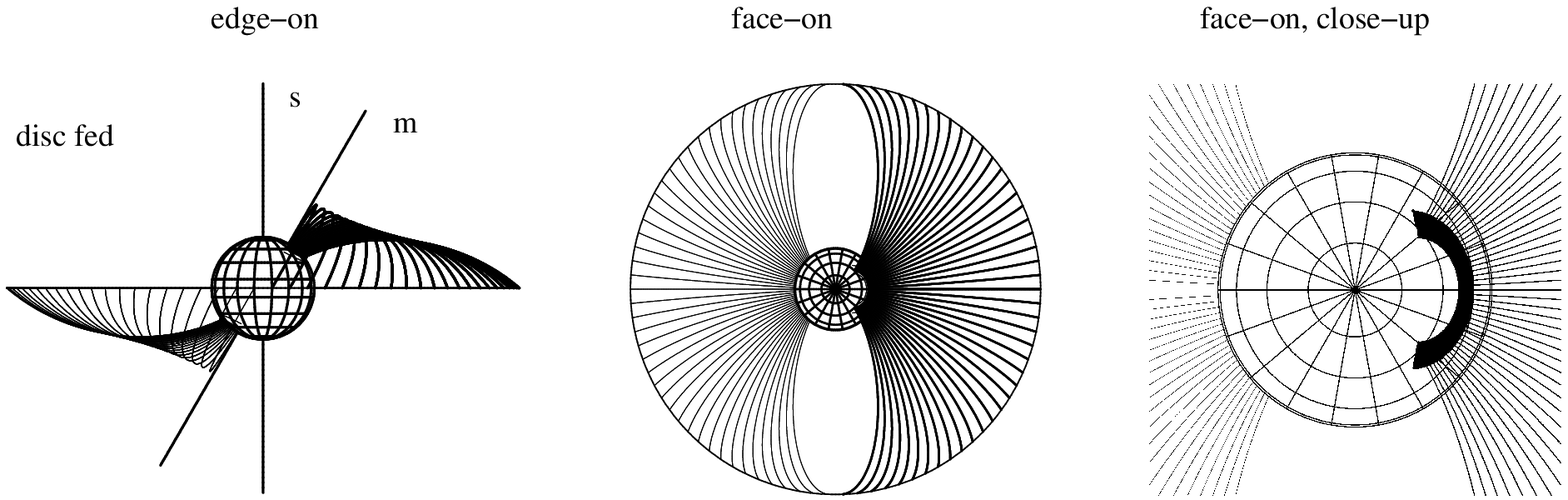}}
\end{picture} 
\end{figure*}
\begin{figure*}
\setlength{\unitlength}{1cm}
\begin{picture}(10,5)
\put(0,0){\includegraphics{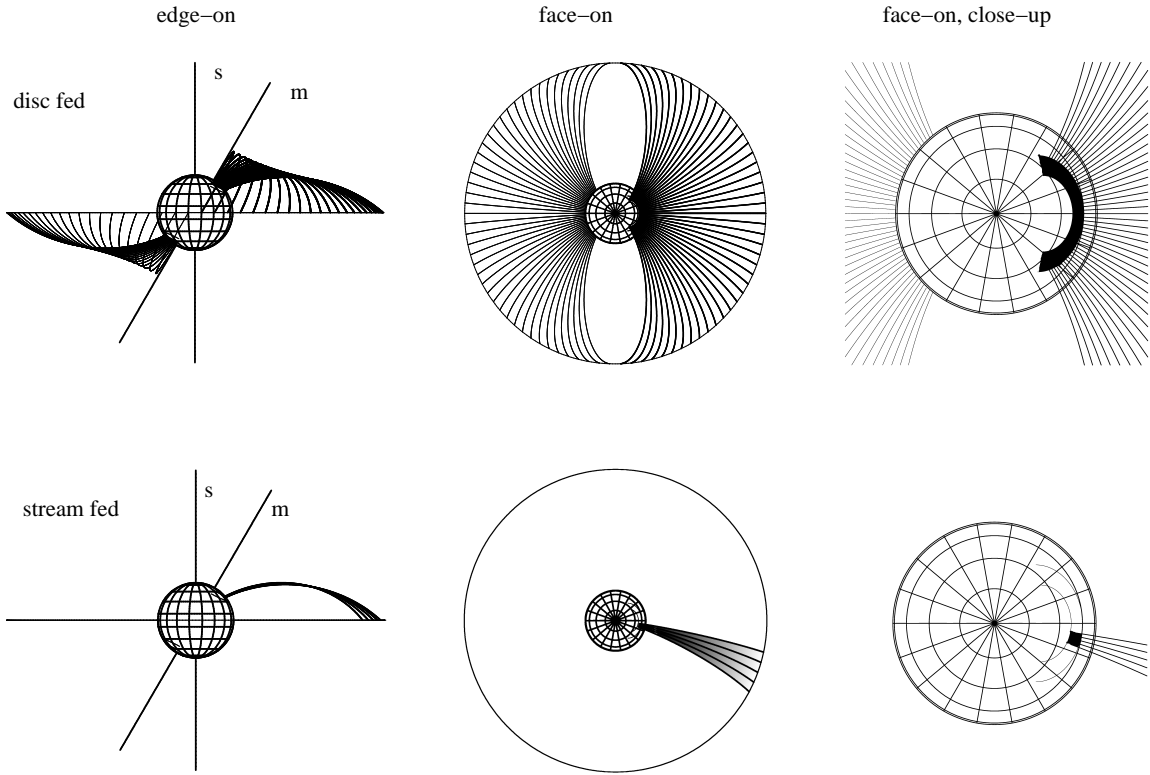}}
\end{picture}
\caption{Schematics illustrating the azimuthal extent of the accretion region
around the upper magnetic pole on the surface of the white dwarf in an IP. 
Upper panels: disc fed accretion; lower panels: stream-fed accretion.  Left
hand panels show the accretion flow and magnetic field lines from the
magnetospheric radius to the white dwarf, viewed edge-on ($i=90^{\circ}$). The
orbital plane, spin axis (s) and magnetic axis (m) of the white dwarf are
indicated; the magnetic axis is inclined at $30^{\circ}$ to the spin axis in
this example. Central panels show the same systems viewed face-on
($i=0^{\circ}$).  The outer circle indicates the magnetospheric radius, set at
five white dwarf radii here. Right hand panels show a close-up of the white
dwarf surface, seen face-on, with the footprint of the accretion flow
highlighted. The footprint of the flow in a stream-fed system, here shown with
an azimuthal extent of $\sim 15^{\circ}$, will migrate around the white dwarf
surface between the two arcs indicated.} 
\end{figure*} 

The accretion flows in IPs have been modelled as a collection of diamagnetic
blobs (e.g. King \& Wynn 1999; Wynn 2000). Those blobs with a high specific
orbital energy will be expelled centrifugally by the white dwarf, whilst lower
energy blobs will be accreted, spinning up the white dwarf. Eventually, the
white dwarf will be spinning so fast that further accretion is prevented and
blobs will be ejected to be swept up by the secondary, spinning the white dwarf
down. Consequently an equilibrium situation results when the rate at which
angular momentum is accreted by the white dwarf is balanced by the braking
effect of the magnetic torque.  This results in an accretion flow which
attaches to the magnetic field at around the co-rotation radius, $r_{\rm co}$
(i.e. the radius at which the magnetic field rotates at the same rate as the
local Keplerian frequency).  

Whether a disc-like structure forms, truncated at $r_{\rm co}$, or whether
accretion is directly via a stream, depends on the relative size of the
magnetospheric radius ($\sim$ the Alfv\'{en} radius, $r_{\rm a}$), which
depends on the white dwarf magnetic moment, and the circularisation radius
$r_{\rm circ}$ at which a circular Keplerian orbit would be established by the
stream.  In a purely stream-fed IP, the white dwarf has a relatively large
magnetic moment and hence $r_{\rm a} > r_{\rm circ}$.  This prevents a disc
forming and material from the stream latches onto the field lines directly.
This will initially spin-up the white dwarf so its corotation radius will
reduce until equilibrium is eventually reached with $r_{\rm a} \sim r_{\rm co}
> r_{\rm circ}$.

The requirement that  $r_{\rm co} > r_{\rm circ}$ leads to a condition
on the spin period of the white dwarf (Wynn \& King 1995)
\begin{equation}
P_{\rm spin} > P_{\rm orb} (0.500 - 0.227\log_{10}q)^6(1+q)^2
\end{equation}
where $q = M_2 / M_1$ is the mass ratio of the system. Since the spin periods
of both RXJ1914 and RXJ0806 are relatively short, this implies that the orbital
periods must be relatively short too. In the limiting case for stability
against thermal timescale mass transfer ($q \sim 1$), $P_{\rm beat} = 569$s
implies $P_{\rm orb} < 2.37$~h for RXJ1914 and $P_{\rm beat} = 321$s implies
$P_{\rm orb} < 1.33$~h for RXJ0806. If $q$ is smaller than this, the upper
limit on the orbital period of each system is even shorter. This indicates that
for RXJ1914 and RXJ0806 to be stream-fed they must have orbital periods below
the period gap and a reasonably high mass ratio.

\subsection{X-ray modulation}

In a stream-fed IP, assuming a dipole magnetic field geometry, the magnetic
field lines from the inclined dipole will intercept the incoming accretion
stream with a varying aspect angle as the white dwarf rotates. After locking on
to the field lines, the accretion flow will preferentially follow the
`downhill' direction to the nearest magnetic pole and the accretion stream will
flip from one pole to the other twice per rotation of the white dwarf (e.g.
Norton 1993).  The upper (visible) pole accretes for half the rotation cycle of
the white dwarf, whilst the lower (hidden) pole accretes for the other half,
giving rise to an X-ray modulation on the beat period. As an example, we show
in Figure 2 a simulation of stream-fed accretion flow in a system with
parameters appropriate to RXJ1914 or RXJ0806, following the prescription of
King \& Wynn (1999) and Wynn (2000). We emphasise that this is merely
illustrative to demonstrate that a stream-fed accretion flow is possible for
the system parameters applicable to these objects.

\begin{figure*}[t]
\setlength{\unitlength}{1cm}
\begin{picture}(10,11)
\put(0,12){\includegraphics{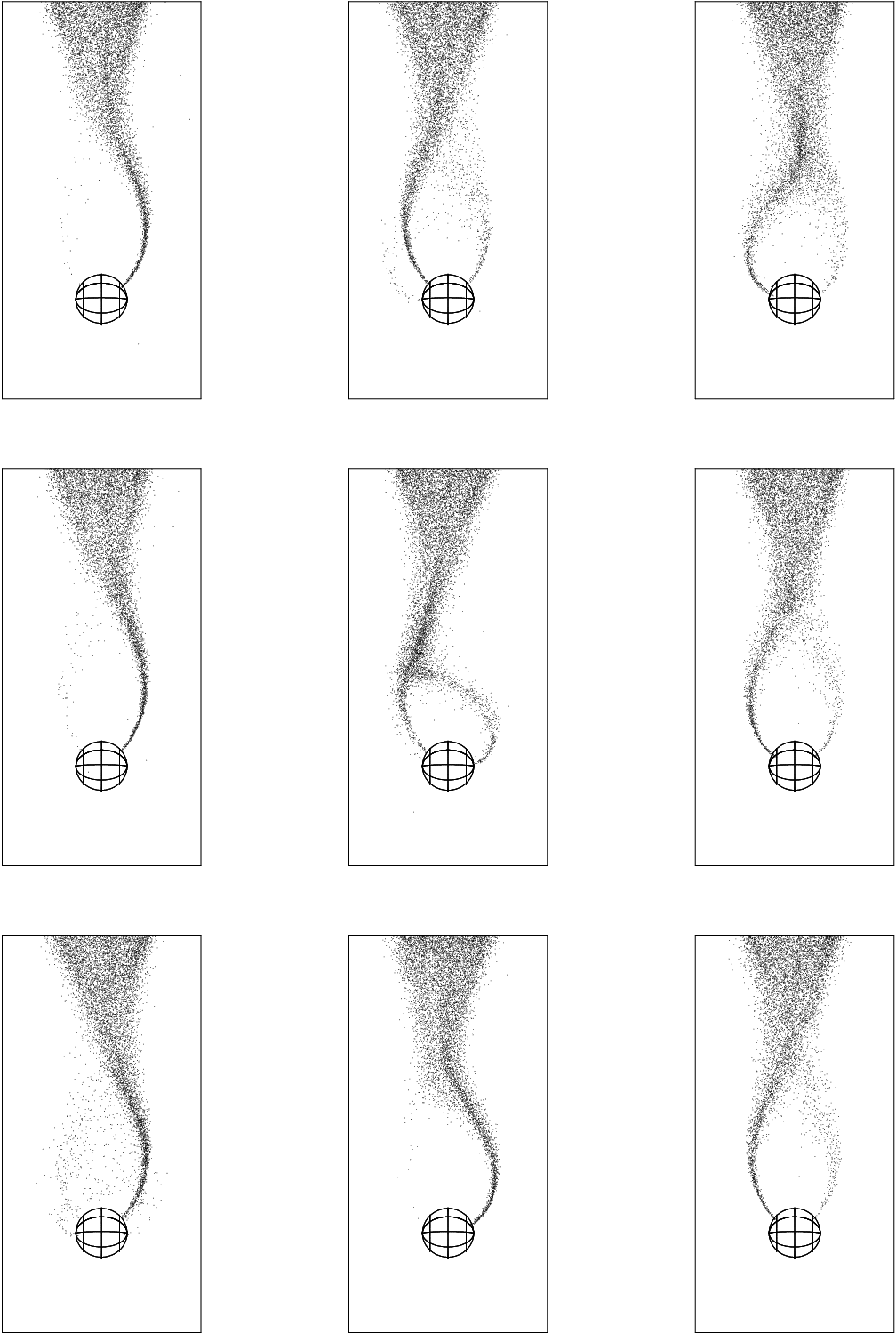}}
\end{picture}
\caption{Stream-fed accretion simulations for a system with 
$P_{\rm orb} = 1.5$~h. The system is here shown edge-on ($i=90^{\circ}$) and
rotated to $20^{\circ}$ after eclipse centre.  The panels span a 
single beat cycle and are equally spaced in beat phase, running from top 
left to bottom right. With reference to the phasing shown in Figure 3,
phase zero occurs around the time of the central panel. Note that in this 
simulation the lower pole accretes preferentially and is only rarely free 
from accretion. This is entirely dependent on initial conditions, as the 
preference for one pole or the other depends on the phase and angle at which 
the stream and field first come into contact at the magnetosphere.} 
\end{figure*}   

The lack of any other X-ray periods is understandable in this model. 
There will be no modulation at the spin period of the white dwarf since there
is nothing which has a structure or a visibility which varies at this period.
We note that Reinsch (2003) reports there is a possible $\sim 4700$~s period
present in both X-ray and optical data from RXJ0806, and that Strohmayer 
detects a sideband structure in the X-ray power spectrum of RXJ1914 indicating
a period close to $\sim 1$ hour. These could represent the orbital periods of 
the systems and their values are consistent with our prediction that 
$P_{\rm orb} < 1.33$~h for RXJ0806 and $P_{\rm orb} < 2.37$~h for RXJ1914.

Cropper et al (1998) rejected the face-on, stream-fed IP model for RXJ1914 on
the grounds that the X-ray emission from the upper pole would not drop to zero
when the stream feeds the lower pole, because the travel time of the accreting
material is of the same order as the 569s period, and also because they thought
it unlikely pole-switching can produce the required sharp rise times seen in
the X-ray pulse profile. Our simulations indicate that the pole-switching is
complete and rapid (Figure 3), with rise and fall times for the number of
particles accreted of order $\sim 0.1$ in beat phase. Thus when the system is
viewed face-on, the X-ray lightcurve will also be fully modulated with a sharp
rise and fall. We recognise that the profile in Figure 3 does not exactly mimic
the X-ray pulse profiles of RXJ1914 and RXJ0806, but emphasise this merely
shows an illustrative accretion duty cycle. Precisely what the X-ray flux duty
cycle would be depends on the details of the shock front at the impact site,
the radiative transfer process and other considerations beyond this simple
model. Nonetheless, it is apparent that a face-on, stream-fed IP model can
produce an X-ray modulation similar to those observed, and rejection of the
model on these grounds is not justified.

\begin{figure}
\setlength{\unitlength}{1cm}
\begin{picture}(6,6)
\put(0,6){\includegraphics{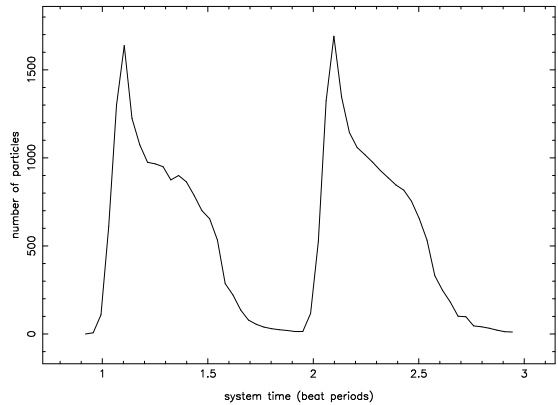}}
\end{picture}
\caption{Results of the stream-fed accretion model illustrated in Figure 2, 
showing the number of `particles' accreted by the upper pole as a function of 
beat phase. For a face-on system, this will be similar to the X-ray 
lightcurve.}
\end{figure} 

The simulations in Figure 2 indicate that the field effectively dams the flow
of matter at certain phases and then allows a burst of accretion as each pole
rotates close to the stream.  This effect is responsible for the sharp spike
seen in Figure 3 at the beginning of each cycle of accretion onto the upper
pole.  The steepness of the rise and fall shown in Figure 3  may be understood
as follows. The free fall time onto a white dwarf from $10^{10}$~cm (the
typical distance at which material might be expected to attach to field lines)
is $\sim 100$~s. However, the time by which the `turn-off' at the upper pole is
delayed as a result of the free-fall time is the same as the delay in the
subsequent `turn-on' at the same pole and a duty cycle of $\sim 50$\% can
result. Hence the rapid rise and fall is essentially independent of the
free-fall time. The rise and fall time {\em does} depend on the azimuthal
extent of the stream. The width of the accretion stream is initially set by
the nozzle at the L1 point (see Equation 5) which in this case is of order
$10^{9}$~cm. The stream width is roughly constant from here to the
magnetospheric radius and will determine the range of azimuth over which
material attaches to the field lines.  Since material then follows the field
lines down to the white dwarf surface, this range in azimuth will be
maintained, although the linear extent of the stream is greatly reduced (Figure
2) as the field lines converge. The azimuthal extent of the arc which forms the
footprint of the accretion flow on the white dwarf surface (Figure 1) will
therefore reflect the original azimuthal extent of the stream. As Figure 3
shows, a stream with initial width of $\sim 10^9$~cm still gives rise to a
sharp rise and fall in the X-ray pulsation.

The rise time of the observed X-ray pulse profiles in both RXJ1914 (Cropper et
al 1998) and RXJ0806 (Burwitz \& Reinsch 2001) extends over $<0.15$ in phase,
indicating an azimuthal stream extent of $<50^{\circ}$. Hellier \& Beardmore
(2002) show that observations of the stream-fed IP V2400~Oph are consistent
with a model in which the azimuthal extent of the accretion stream is around
$20^{\circ}$.

V2400 Oph is the clearest example of a stream-fed IP previously known (Buckley
et al 1995; Buckley et al 1997; Hellier and Beardmore 2002) although FO~Aqr
(Beardmore et al 1998) and TX~Col (Norton et al 1997) have each been seen to
switch between predominantly disc-fed and predominantly stream-fed accretion
modes on timescales of years or months. V2400 Oph is also a low inclination
system with $i < 10^{\circ}$ (Buckley et al 1997). In this case, the X-ray
pulse is roughly sinusoidal and, unlike the case of RXJ1914 and RXJ0806, does
not go to zero flux for half the cycle. In order to explain this modulation,
Hellier \& Beardmore (2002) were forced to postulate the existence of a ring of
material around the white dwarf which accretes continuously onto both poles. It
is only by invoking this non-stream element to the accretion that they were
able to justify the fact the the X-ray flux in V2400 Oph does not drop to zero.

\subsection{X-ray spectra}

Both RXJ1914 and RXJ0806 have X-ray spectra that are well fitted by simple
blackbody models with temperatures of 43 eV (Motch et al 1996; Cropper et al
1998) and 64 eV (Israel et al 2003a) respectively. Neither object shows
evidence for the hard X-ray component seen in other IPs, and on these grounds
Israel et al (2003a) rejected the IP interpretation of RXJ0806. However, unlike
other IPs, the X-ray emitting regions in RXJ1914 and RXJ0806 are fuelled purely
by a stream-fed accretion flow. This means that the accretion footprint is much
smaller than in the case of disc-fed systems. We also suggest that both systems
have relatively high accretion rates (see Section 2.9). Given these constraints
it is more likely that the accreting material becomes buried beneath the white
dwarf surface before releasing X-rays (Frank, King \& Lasota 1988). The outcome
of such an accretion mode is a blackbody X-ray spectrum, as observed in these
systems. In the other stream-fed system, V2400 Oph, part of the flow is
disc-fed (Hellier \& Beardmore 2002) and so will have a conventional flow,
producing hard X-ray bremsstrahlung emission from beneath a shock situated 
above the white dwarf surface.

\subsection{Optical and infrared modulation}

The azimuthally concentrated footprint of the accretion stream on the white
dwarf surface will migrate around the magnetic pole to follow the incoming
stream (Figure 1), as the field lines which have captured the flow change on
the synodic spin period (Norton 1993).  We suggest this will leave behind a
heated trail on the white dwarf surface which will be a source of
optical/infrared blackbody radiation.  The upper magnetic pole of the white
dwarf will be heated by the accretion flow for half the rotation period.  When
the stream flips to the lower pole, the heated trail will be completely exposed
and the optical to infrared emission from the whole of this heated region will
be seen. As long as some part of the optical to infrared emission from the
heated trail is absorbed by the flow whilst the accretion stream is feeding the
upper pole, then a residual optical to infrared modulation will remain and be
antiphased with the X-ray.

To assess the viability of this scenario for RXJ1914, we assume $P_{\rm orb}
\sim 2$~h, so the secondary star has $M_2 \sim 0.2$~M$_{\odot}$, $R_2 \sim
0.2$~R$_{\odot}$ and $T_2 \sim 3000$~K. In order to have $r_{\rm co} > r_{\rm
circ}$ (Equation 2), this orbital period requires $q \sim 0.6$, so we assume
for the white dwarf, $M_1 \sim 0.3$~M$_{\odot}$ and therefore $R_1 \sim 1.25
\times 10^{9}$~cm. For illustrative purposes we further assume that the heated
trail left behind after the stream has flipped to the lower pole has a
temperature of order $10^5$~K and occupies an area on the white dwarf of order
0.2\% of its surface. This area is $\sim$ a few times the hotspot area in
polars (Warner 1995) since it is the {\em trail} of a stream we must consider
here, not the stream footprint area itself. We then assume that when the stream
is feeding the upper pole, one-fifth of the heated trail is obscured by the
stream itself. Then, in the V-band, the ratio of blackbody flux from the
obscured part of the heated trail to that from the secondary star plus
contributions from elsewhere in the system (assumed to be comparable to that
from the secondary star) is of order 15\%, in agreement with the $\sim 15\%$
modulation observed in the V-band flux (Ramsay et al 2002a). A similar 
calculation and conclusion applies to RXJ0806, although here the secondary
star is likely to be even fainter and so contribute even less light (Reinsch
2003).

The X-ray spectrum of RXJ1914 is best fit by a blackbody of temperature
43~eV (Motch et al 1996, Cropper et al 1998), corresponding to $T \sim 5
\times 10^5$~K. We assume this to be the temperature of the stream footprint,
with an area $\sim 0.04\%$ of the white dwarf surface, whilst material is
feeding the visible pole. The ratio of the X-ray flux in the {\em ROSAT} band
(0.1 -- 2 keV) from blackbodies at temperatures of $5 \times 10^5$~K and
$10^5$~K, with the cooler one occupying five times greater area, is about 150.
Therefore the heated trail left behind after the stream has flipped to the
lower pole would indeed give rise to an X-ray modulation with an amplitude of
close to 100\%, as observed (Cropper et al 1998). Once again, a similar result 
may be obtained for RXJ0806.

\subsection{Optical spectra}

The optical spectrum of RXJ0806 shows faint, broad emission lines which are
claimed to be mostly those of the HeII Pickering series (Israel et al 2002),
i.e. transitions from $n = 5, 6, 7, 8 ... $ to 4. However, the lines
corresponding to even terms of the series are all stronger than those of the
odd terms. The most likely explanation for this is that the even lines are each
blended with those of the Balmer series which occur within 2\AA \ in each case
(HeII Pickering $\beta$ $\sim$ H$\alpha$, HeII Pickering $\delta$ $\sim$
H$\beta$, etc). This is unlike the spectra of any of the known AM CVn stars
(e.g. Marsh et al 1991; Groot et al 2001; Ruiz et al 2001). None of these
interacting double degenerate systems show any evidence for hydrogen and none
of them show the HeII Pickering series.  Any hydrogen in the precursor to an
AM CVn system is believed to be lost during the two common envelope phases that
are necessary to achieve the close orbit.  If RXJ0806 is a double degenerate
with an orbital period of 321~s, it too should have lost all its hydrogen.  The
shortest orbital period possible for a system with a degenerate hydrogen-rich
donor star is $\sim 30$ minutes (Rappaport, Joss \& Webbink 1982). Hence the
presence of hydrogen in the spectrum of RXJ0806 argues against the three double
degenerate binary models for this system. 

Similar blending of Balmer lines with HeII lines has been seen in the outburst
spectra of the soft X-ray transient XTE J2123--058 (Hynes et al 2001). In
quiescence XTE J2123--058 has a well-studied late K dwarf absorption spectrum
with H$\alpha$ emission from the accretion disc (Casares et al 2002, Tomsick et
al 2002); no abundance anomalies have been reported. Hence normal abundance
material can produce a spectrum resembling that of RX\-J0806 if the physical
conditions are appropriate. In section 2.3 we showed that the stream-fed IP
model predicts a small accretion footprint and high mass transfer rate, hence
at the impact site on the white dwarf these systems are expected to have more
extreme physical conditions than other IPs, and indeed the resemblance of the
RXJ0806 spectrum to that of the accreting neutron star in XTE J2123--058 is not
unexpected.  

As noted earlier, the fact that the optical spectrum of RXJ1914 has
spectral features that are consistent with those of a K star (Steeghs, 
private communication) poses a question for the double degenerate models for 
that system too.

A remaining issue to consider is the lack of emission lines in the optical and
infrared spectra of RXJ1914 (Ramsay et al 2002a). In conventional (disc-fed)
CVs, the emission lines are presumed to originate in the accretion disc itself,
from the bright spot where a stream impacts the disc, from a corona above the
disc, or (in the case of some UV lines) from a wind emanating from the system
(Warner 1995).  Clearly, a stream-fed IP would not be expected to exhibit
emission lines from such locations. However, the disc-less polars also exhibit
emission lines, and here the line emission is attributed to the accretion
stream itself. A similar origin for emission lines might be expected in
stream-fed IPs.  The majority of the line emission in polars arises near the
base of the accretion stream where the stream material is photoionized by the
UV and X-ray flux from below the shock (Warner 1995). For the accretion stream
in RXJ1914, we have already suggested that the stream material has a
significant optical depth to the optical/infrared continuum from the heated
trail on the white dwarf surface, so it will have an even greater optical depth
to optical/infrared emission lines.  

To estimate the optical depth of the stream, we first calculate its optical
depth as it emerges through the L1 point. Here, the width of the stream may 
be approximated by
\begin{equation}
W \sim c_s P_{\rm orb} / 2 \pi
\end{equation}
and the density at this point is given by
\begin{equation}
\rho \sim \dot{M} / c_s W^2 .
\end{equation}
So assuming $T \sim 3000$~K, $P_{\rm orb} \sim 2$~h, the sound speed 
$c_s \sim 10^6$~cm~s$^{-1}$ and the mass transfer rate $\dot{M} \sim 
10^{17}$~g~s$^{-1}$ (see Section 2.9), we have $W \sim 10^9$~cm and $\rho \sim 
10^{-7}$~g~cm$^{-3}$. Using the grid of low temperature opacities 
presented by Alexander \& Ferguson (1994), this combination of temperature
and density corresponds to an opacity at the L1 point of $\kappa \sim 
0.03$~cm$^2$~g$^{-1}$. Hence the optical depth at the L1 point, given by 
\begin{equation}
\tau = \kappa \rho W
\end{equation}
is $\tau \sim 3$. Since the stream is essentially confined from here down to
the white dwarf surface, it will become more concentrated as the field lines
converge near to the magnetic poles. At the white dwarf surface, the width of
the stream is $W \sim 10^7$~cm and the density will therefore increase to $\rho
\sim 10^{-3}$~g~cm$^{-3}$. Assuming that the opacity obeys $\kappa \propto
\rho^{1/2}$, the opacity will therefore be $\sim 100 \times$ higher at the base
of the stream (for constant temperature along the stream). The optical depth at
this point is then $\tau \sim 3 \times 10^4$, and the base of the accretion
stream will indeed be extremely optically thick. 

Allowing for the fact that the base of the accretion stream near to the white
dwarf will be hotter than at the L1 point, the opacity will not be less than
$\kappa \sim 0.3$~cm$^2$~g$^{-1}$ which is the limit imposed by electron
scattering at high temperatures. Even this opacity corresponds to $\tau \sim 3
\times 10^3$ for the density and width of the stream at this point.
Furthermore, in a face-on system with a magnetic axis angle of a few 
tens of degrees, this region will always be viewed through
parts of the stream that are further out, and hence cooler and more optically
thick. So it is unlikely that any emission lines will be seen from a
stream-fed, face-on IP with a high mass transfer rate.

\subsection{Radial velocity variations}

As noted above, even though emission lines are suppressed in RXJ1914, the
accretion stream in a stream-fed IP may be a source of optical
emission lines. Close to the white dwarf where these are emitted, the velocity
of the material is likely to be of order $\sim 1000$~km~s$^{-1}$, and unless
the system is seen precisely at an inclination of $0^{\circ}$, radial velocity
shifts of these lines would be observable and vary on the orbital period of the
system. However, the current data on the lines in RXJ0806 are too poor to
reveal such motions, and no emission lines are seen in RXJ1914 anyway.  This
suggests a test of the IP model, in that detailed radial velocity studies of
the emission lines in RXJ0806 should reveal a longer orbital period modulation.

\subsection{Polarization}

The lack of polarization seen in RXJ1914 (Ramsay et al 2002a) is consistent
with the IP interpretation, as the majority of IPs have lower magnetic field
strengths than polars and do not exhibit polarized emission.  Only 5 out of
$\sim 25$ confirmed IPs have been seen to emit polarized light.  Similarly, the
recent detection of linear polarization at the $1.7\%$ level in RXJ0806 (Israel
et al 2003b) is in agreement with the level of polarization seen in the few IPs
that do reveal polarized emission. However, this low level and its 
lack of variation is also consistent with the polarization being interstellar 
in origin.

\subsection{Period change}

As noted above, IPs are expected to evolve towards an equilibrium spin period
with $r_{\rm co} \sim r_{\rm a}$ and a continuum of equilibrium spin rates
exists with spin periods varying as a function of orbital period and white dwarf
magnetic moment (Wynn 2000; Norton, Somerscales \& Wynn 2003). However, on
short timescales ($\sim$ years) random spin-up or spin-down episodes may be
expected due to fluctuations in the mass transfer rate.  Amongst other IPs,
some are seen to be spinning down and others to be spinning up (e.g. Patterson
1994), with the $\dot{P}$ of FO~Aqr having been observed to change sign over
the last few years.

If the periods in RXJ1914 and RXJ0806 arise from rotating magnetic white
dwarfs, then the recent measurements by Strohmayer (2002; 2003; 2004) and 
Hakala et al (2003) imply that the white dwarfs in each system are spinning up. 
(Although, as already noted, Woudt \& Warner (2003) suggest that the period
derivative in RXJ0806 is ambiguous.) Even if the period derivatives are
confirmed, the few year span over which these measurements were obtained does
not necessarily reflect a secular period derivative of the magnetic white
dwarf. The reported rate of change of frequency in RXJ1914 corresponds to
$\dot{P} \sim 2 \times 10^{-12}$~s~s$^{-1}$ (Strohmeyer 2004) whilst that in 
RXJ0806 corresponds to $\dot{P} \sim 6 \times 10^{-11}$~s~s$^{-1}$ (Hakala et 
al 2003; Stromayer 2003). These are comparable with those
seen in other IPs (Patterson 1994) and the spin up timescales are $P_{\rm
spin}/\dot{P} \sim 9 \times 10^6$~years and $\sim 2 \times 10^5$~years
respectively. These are typical for a white dwarf and therefore are consistent
with an IP model. It is entirely feasible that the period changes represent
temporary spin-up episodes, and the systems are accreting via a stream with a
long term white dwarf spin rate equal to its equilibrium value. Such a
temporary spin-up phase may be the result of a large mass accretion rate which
in turn is due to magnetic activity on the secondary. The magnetic field of the
secondary may dominate the flow near the L1 point if the region of the
secondary star near the L1 point is magnetically active (Barrett et al 1988).

\subsection{Mass accretion rate}

The measured spin-up rate may be used to estimate the mass accretion
rate in RXJ1914. As above we assume $P_{\rm orb} \sim 2$~h, $M_2 \sim 
0.2$~M$_{\odot}$, $M_1 \sim 0.3$~M$_{\odot}$ and $R_1 \sim 1.25 \times 
10^{9}$~cm. The orbital separation of the two stars is therefore 
$a \sim 4.4 \times 10^{10}$~cm and the distance from the white dwarf to the 
L1 point is 
\begin{equation}
b = a(0.500 - 0.227 \log_{10} q) \sim 2.4 \times 10^{10}~{\rm cm}
\end{equation}
The moment of inertia of the white dwarf is
\begin{equation}
I = \frac{2}{5}M_1 R_1^2 \sim 3.75 \times 10^{50}$g~cm$^{2}
\end{equation}
With a frequency derivative of 
$\dot{f} = 6 \times 10^{-18}$~Hz~s$^{-1}$ (Strohmayer 2004), the 
implied mass transfer rate is
\begin{equation}
\dot{M} = \frac {\dot{f} I P_{\rm orb}}{b^2} \sim 3 \times 
10^{16}~{\rm g~s}^{-1} \sim  5 \times 10^{-10} {\rm M}_{\odot}~{\rm yr}^{-1}
\end{equation}
This may be compared with the mass accretion rate derived from the X-ray
spectral properties. RXJ1914 has an X-ray spectrum best fit by a blackbody with
a temperature of 43~eV (Motch et al 1996, Cropper et al 1998). The emitted
blackbody flux is therefore $F_{\rm BB} = 3.5 \times
10^{18}$~erg~cm$^{-2}$~s$^{-1}$.  If we assume the accretion stream impacts the
white dwarf over an area of $\sim 0.01\%$ of its surface area as in polars
(Warner 1995), the blackbody accretion luminosity is $L_{\rm BB} \sim 7 \times
10^{33}$~erg~s$^{-1}$ which corresponds to an accretion rate $\dot{M} \sim 2
\times 10^{17}~{\rm g~s}^{-1} \sim 3\times 10^{-9} {\rm M}_{\odot}~{\rm
yr}^{-1}$ in rough agreement with the estimate from the spin-up rate.  

These mass accretion rates are relatively high for an IP and support the
suggestion that a dense accretion stream is responsible for suppressing
the line emission in RXJ1914. We note that the density of the stream is
enhanced over that in a conventional disc-fed IP with the same $\dot{M}$, since
the flow in the stream-fed case is concentrated in a region which extends over
only a few tens of degrees in azimuth around the magnetic pole compared with
the $180^{\circ}$ extent of an accretion curtain in a disc-fed IP (Figure 1).

\subsection{Probability of detection}

Of the four models proposed to explain RXJ1914 and RXJ0806, three have
potential difficulties relating to how likely it is that such systems are
detectable. In the double degenerate Algol model the parameter space imposed by
the geometric constraints on the system is rather confined (Marsh \& Steeghs
2002; Ramsay et al 2002b). In the electric star model the predicted lifetime of
the phase is only $\sim 10^3$ years (Wu et al 2002). However, the system
can undergo many cycles of this behaviour and so can appear as an electric
star at many epochs (Cropper, Ramsay \& Wu 2003). This is therefore not a 
serious difficulty for the model.

The apparent problem with the face-on, stream-fed IP model is that only $\sim
1.5\%$ of all IPs would be expected to have an inclination angle $i <
10^{\circ}$. Since V2400 Oph is already suggested to be such a system, further
examples would not necessarily be expected amongst the 25 or so confirmed IPs
in the absence of selection effects.  However, it may be that the constraint on
inclination angle necessary to produce a fully modulated X-ray pulse profile
can be relaxed somewhat. King \& Shaviv (1984) and Wynn \& King (1992) show
that, based on geometric considerations only, the upper magnetic pole remains
constantly visible, and the lower pole is never seen, as long as:
\begin{equation}
i + m < 90^{\circ} - \beta
\end{equation}
where $m$ is the angle between the dipole magnetic axis and the spin axis of
the white dwarf and $\beta$ is the angle subtended at the magnetic axis by the
X-ray emitting region. As noted earlier, $m$ is typically $\sim 10^{\circ} -
30^{\circ}$, and $\beta$ will be similar. Based on this simple geometry,
inclination angles of up to several tens of degrees would still give rise to a
fully modulated X-ray pulse in the case of stream-fed accretion.

The fact that all three IPs with a substantial stream-fed component are close
to face-on may not be unexpected though. In a stream-fed IP seen at high
inclination, both poles would be visible at some times during the beat cycle,
and little modulation would occur as the disappearance of one pole is
compensated by the reappearance of the other. Therefore there is an
observational bias against high inclination stream-fed systems and in favour of
low inclination stream-fed systems.

The fact that all three face-on IPs have a stream-fed component is not
surprising either, since a purely disc-fed face-on IP would display no X-ray
modulation. The structure and visibility of the X-ray emitting upper pole would
remain constant throughout the spin cycle of the white dwarf in such a case. 
Therefore any face-on disc-fed IPs would almost certainly go un-recognised
because X-ray variability is the usual key to the identification of an IP.  

In conclusion, the only purely stream-fed IPs that will display an X-ray
modulation are those that are seen close to face-on, and the only face-on IPs
that will display an X-ray modulation are those in which at least some of the
accretion arrives at the white dwarf without first flowing through a disc.

\section{Conclusions and predictions}
 
It is difficult to see how the double degenerate models are viable given
the observation of hydrogen lines in the spectrum of RXJ0806 (Israel et al
2002) and the spectral features consistent with those of a K star 
seen in RXJ1914 (Steeghs, private communication).
If the period decreases seen in RXJ1914 (Strohmayer 2002) and RXJ0806
(Stroh\-meyer 2003; Hakala et al 2003) are confirmed, and do indeed represent
secular changes in the period, this too would rule out accretion from a
degenerate donor star and the identification of the periods as orbital in
nature. We note though that such period changes may also be understood in terms
of magnetic cycles on the donor star, so this evidence is not compelling. The
lack of polarization in RXJ1914 and low level polarization in RXJ0806 are a
problem for the double degenerate polar model, whilst the geometrical
constraints argue against the double degenerate Algol model.

For both RXJ1914 and RXJ0806, the X-ray pulse profiles, X-ray spectra, lack of
other modulation periods, antiphased optical/infrared modulation, and level of
polarization may be understood in terms of a face-on, stream-fed intermediate
polar model. The implied orbital periods in each case are below the period gap
(RXJ1914: $P_{\rm orb} < 2.37$~h; RXJ0806: $P_{\rm orb} < 1.33$~h), and the
mass ratios ($M_2/M_1$) are relatively high. In the case of RXJ1914, the lack
of spectral lines and the observed rate of period decrease may all be
attributed to a current phase of high mass accretion rate, with $\dot{M} \sim
10^{17}$~g~s$^{-1}$. The possible detection of sideband structure in the 
X-ray power spectrum of RXJ1914 (Strohmayer 2004) and the possible long
period X-ray modulation in RXJ0806 (Reinsch 2003) each suggest that the 
originally detected short period is related to a white dwarf spin period
rather than an orbital period. Finally, the preferential detection of 
stream-fed systems with a low inclination angle may be the result of 
observational selection effects.

If the face-on stream-fed IP model is correct, then the emission lines seen in
RXJ0806, and any which are ever detected from RXJ1914, should exhibit no
sinusoidal radial velocity variations at the 321s or 569s periods, but may show
modulation at a longer orbital period. If, instead, 321s and 569s do represent
orbital periods in double degenerate binaries then sinusoidal radial velocity
variations should be present at these periods.  In the face-on, stream-fed IP
model, there may be a kinematic signature of the flipping accretion stream in
any lines which are found. For half the beat cycle, the stream flows
essentially towards the observer before turning to crash onto the upper,
facing, magnetic pole of the white dwarf. For the other half of the beat cycle,
the stream flows away from the observer, before turning to crash onto the
lower, hidden, magnetic pole.  This stream-flipping may lead to velocity
variations of spectral lines. Any variation would be at the 569s or 321s period
in the two systems but would not be a simple sinusoidal modulation.

\begin{acknowledgements}
We thank Andy Beardmore for the software used to produce Figure 1 and Matt
Burleigh for useful discussions. We also thank the referee for several
useful suggestions, including the possibility of a double degenerate IP.
\end{acknowledgements}

\end{document}